\documentstyle[preprint,aps,prl,epsf]{revtex}
\begin{document}
\def\alt{\mathrel{\mathpalette\vereq<}}
\def\agt{\mathrel{\mathpalette\vereq>}}
\def\e{{$\grave{e}$}}
\def\ep{{\epsilon}}
\def\tg{{\tilde{\cal G}}}
\renewcommand{\thefootnote} {\fnsymbol{footnote}}

\draft
\title{``Exhaustion'' Physics in the Periodic Anderson Model from
Iterated Perturbation Theory}
\author{N. S. Vidhyadhiraja$^{1,2}$,A. N. Tahvildar-Zadeh$^{1}$,
M. Jarrell$^{2}$, and H. R. Krishnamurthy$^{2}$}
\address{$^1$ Department of Physics, University of Cincinnati,
Cincinnati,
OH 45221}
\address{$^2$ Department of Physics, IISc, Bangalore 560012, India}
\date{\today}

\maketitle

\begin{abstract}
We discuss the ``exhaustion'' problem in the context of the Periodic
Anderson Model using Iterated Perturbation theory within the Dynamical
Mean Field Theory.  We find that, despite its limitations, IPT captures 
the exhaustion physics, which manifests itself as a dramatic, strongly
energy dependent, suppression of the effective hybridization of the
self consistent Anderson impurity problem. As a consequence, low energy 
scales in the lattice case are strongly suppressed compared to the 
``Kondo scale'' in the single impurity picture.The IPT results are in 
qualitative agreement with recent Quantum Monte Carlo results for the 
same problem.
\end{abstract}


\section{Introduction}
Metallic compounds containing rare earth elements with partially
filled $f$ shells, such as CeBe$_{13}$ or UPt$_{3}$, belong to
the general category of heavy Fermion materials\cite{review}. They  are
characterized by large Pauli susceptibility and  specific
heat coefficient as compared to ordinary metals, which indicate a huge
effective electronic mass;  and also by anomalous transport
properties such as  non-monotonic temperature dependence of the
resistivity. These anomalies are usually attributed to the
formation of a narrow resonant state at the Fermi energy due to the
admixture of strongly correlated local $f$
orbitals with the metallic band of the host. 
Such materials are normally modelled in terms of
an asymmetric Periodic Anderson Model(PAM). But many of the high 
temperature properties of these materials are surprisingly similar
to those of a Single Impurity Anderson Model(SIAM)\cite{review2};
so the formation of
this resonance has also been sometimes interpreted in terms of the
SIAM~\cite{photo}.  However, an impurity treatment clearly neglects
the effects of correlations between the impurity sites and also of
lattice coherence (i.e. Bloch's theorem) which leads to what 
Nozi\e res~\cite{noz1,noz2} termed as ``exhaustion''.

The ``Exhaustion'' problem~\cite{noz1,noz2}, originally posed by 
Nozi\e res in the context of a {\em Kondo lattice },
occurs when a {\it few} mobile electrons, $N_{scr}$,
have to coherently screen {\it many} local moments, $N_f$, in a metallic
environment, i.e, $N_{scr} << N_f$. 
This situation is engendered by the fact that only the
electrons roughly within $T_K$ (the single 
impurity Kondo temperature) of the Fermi surface can 
effectively participate in screening the local moments.
Thus the  number of screening electrons can be estimated as
 $N_{scr}=\rho_d(\ep_F) T_K$,
where $\rho_d(\ep)$ is the {\em conduction band} density of 
states(DOS) and $\ep_F$ is the Fermi level. 
A measure of exhaustion is the dimensionless ratio~\cite{noz1}
\begin{equation}
p=\frac{N_f}{N_{scr}}=\frac{N_f}{\rho_d(\ep_F) T_K}\,.
\label{noztkp}
\end{equation}
Nozi\e res has argued that $p$ is roughly the number of
scattering events between a local moment and
a mobile electron necessary for the mobile electron's spin to precess
around a local moment by $2\pi$\cite{noz2}. In the
case $p\gg 1$, when the number of screening electrons is much smaller
than
the number of local moments to screen, magnetic screening is necessarily
collective and the single impurity picture becomes invalid.

In a recent study\cite{niki} of the 
PAM using the Dynamical Mean Field Theory(DMFT), which is exact 
in the $\infty-$dimensional limit\cite{metzvoll}, it was argued
that exhaustion was responsible for the severe reduction of the Kondo scale
of the PAM from that of the impurity problem with the same
parameters. The issue was  explored in detail\cite{niki} using
the Quantum Monte Carlo--Maximum Entropy Method(QMC-MEM) for solving 
the self--consistent Anderson Impurity problem of the DMFT.  The
reduction of the Kondo scale and the crossover between the two scales
was interpreted in terms of Nozi\e res exhaustion principle~\cite{noz1},
and his mapping of the model which has exhausted all available
conduction band
screening states onto an effective Hubbard model. Using qualitative
and semi-quantitative arguments, Nozi\e res~\cite{noz2} has
suggested that the temperature scale $T_c$  associated with the onset of
Fermi liquid coherence in this ``exhaustion'' limit is suppressed compared to
the Kondo temperature of the single-impurity Anderson Model (SIAM), $T_K$,
by the factor $p$
\begin{equation}
T_c\simeq T_K/p = T^2_K \rho_d(\ep_F)/N_f \,.
\label{Tck}
\end{equation}
While the aforementioned suppression of $T_c$ was clearly evident in the
QMC simulations, no universal relation of $T_c$ to $T_K$ such
as Eq.~\ref{Tck} was found.

In this work, we explore the problem of exhaustion in the context of 
the asymmetric PAM once again within the DMFT, but
using Iterated Perturbation theory (IPT)~\cite{kajuet} for solving 
the self--consistent impurity problem . We calculate
the spectral functions and as a check on the accuracy of IPT, compare
them with earlier QMC-MEM results~\cite{niki}. We find
reasonable agreement between the two.  The full width at half 
maximum(FWHM) of the ``Kondo Resonance'' in the spectral functions 
calculated for the PAM and the SIAM provides {\em one measure} of 
the ``coherence temperature'' $T_c$ and the ``Kondo Temperature'' $T_K$ 
respectively.  We find that, consistent with the ``Protracted screening'' 
picture of ref~\cite{niki} and the ``Collective Kondo
screening'' picture of Nozi\e res~\cite{noz2}, a key feature
of exhaustion, namely the suppression of $T_c$ as compared to 
$T_K$, is recovered here as well. 
Furthermore, we find $T_c$ and $T_K$ to be related as
\begin{equation}
T_c=\frac{T_K}{\alpha(U,V)p_0}\,. 
\label{Tckp0}
\end{equation}
Here $\alpha(U,V)$ is the fitting parameter and $p_0$ is defined 
(similarly to Eq~\ref{noztkp})
as $p_0\equiv N_f/(\rho(x)T_K)$, where $\rho(\ep)$ is the bare 
conduction DOS and $x$ is determined by
\begin{math}
n_d=2\int_{-\infty}^x d\ep\, \rho(\ep) \,.
\end{math}


\section{ Model and Formalism}
The PAM Hamiltonian on a $D$-dimensional hypercubic lattice is
\begin{eqnarray}
H&=&\frac{-t^*}{2\sqrt{D}}\sum_{<ij>\sigma}(d^{\dagger}_{i\sigma}
d_{j\sigma} + H.c) +
\sum_{i\sigma}(\ep_dd^{\dagger}_{i\sigma}d_{i\sigma}
+\ep_ff^{\dagger}_{i\sigma}f_{i\sigma})  \nonumber \\
& &+V \sum_{i\sigma}(d^{\dagger}_{i\sigma}f_{i\sigma}+H.c) +
U\sum_{i}n_{fi\uparrow}n_{fi\downarrow} \,.
\label{pammodel}
\end{eqnarray}
In Eq. \ref{pammodel},
$d^{(\dagger)}_{i\sigma}(f^{(\dagger)}_{i\sigma})$
is a creation operator that creates a $d(f)-$electron with spin $\sigma$
on site i; $d_{i\sigma}(f_{i\sigma})$
is the corresponding destruction operator.  The hopping is restricted to
the nearest neighbors and scaled as $t=t^*/2\sqrt{D}$.   $U$ is the
on-site Coulomb repulsion for the localized $f$ states, $V$ is the
hybridization between $d$ and $f$ states and $\ep_f, \ep_d$
are the site energies for $f$ and $d$ electrons.

We work in the $D\rightarrow \infty$ limit where it was shown by Metzner
and Vollhardt \cite{metzvoll} that the irreducible self-energy and the
vertex functions become purely local. As a consequence, the
interacting lattice model can be mapped onto a local correlated
impurity coupled to an effective bath that is self-consistently
determined~\cite{mapping,footn1}.
In this infinite dimensional limit the non--interacting DOS
has the Gaussian form\cite{metzvoll}
\begin{math}
\rho({\ep})=\frac{1}{\sqrt{\pi\,t^*}}\exp\left[-\frac{\ep^2}
{{t^{*}}^2}\right] \,.
\end{math}
We choose our energy scale such that $t^*=1$.

The local $f$ and $d$-propagators for the PAM are given by
\begin{eqnarray}
G_{d,loc}(\omega)&=&\tilde{D}\left(\omega^+-\ep_d -\frac{V^2}
{\alpha(\omega)}\right) \label{gdloc}  \\
\mbox{and } G_{f,loc}(\omega)&=&\frac{1}{\alpha(\omega)}\left[1+\frac{V^2}
{\alpha(\omega)}G_{d,loc}(\omega)\right] \label{gfloc}
\end{eqnarray}
where $\tilde{D}(z)\equiv\int_{-\infty}^{\infty}d\ep\frac{\rho(\ep)}
{z-\ep}$ is the Hilbert Transform of $\rho(\ep)$
and $\alpha(\omega)\equiv\omega^+-\ep_f -\Sigma(\omega^+)$
where $\omega^+=\omega+i0$.

The self--consistent host is determined by the bare local propagator of the
effective single--site problem as
\begin{equation}
{\cal {G}}^{-1}=G_{f,loc}^{-1}+\Sigma  \label{eq:calg}
\end{equation}
and self--consistency is achieved through the use of Eqs.~\ref{gdloc},
\ref{gfloc} and
the result for the self--energy for the single--site, or the impurity
problem $\Sigma\equiv\Sigma({\cal G})$ . 

Given a starting self--energy, we use Eqs.~\ref{gdloc},\ref{gfloc}
and \ref{eq:calg} to compute $\cal{G}$, and then a
prescription for the effective single--impurity problem to calculate the 
new self--energy.  This procedure is repeated until self--consistency is 
achieved.
We use Iterated Perturbation Theory(IPT)~\cite{kajuet} as the
prescription to calculate the self-energy for the effective impurity 
problem. The motivation for using this scheme is that it is semi-analytical 
and much easier to implement than Quantum Monte Carlo(QMC).  While it 
has the disadvantage that it is perturbative,its advantages are that 
we obtain real--frequency data directly at zero temperature and can study
a wide range of parameters with much less effort compared to the
QMC method. We now briefly review this scheme.

The IPT ansatz~\cite{kajuet} for the total self--energy is given by
\begin{equation}
\Sigma_{int}(\omega)=\frac{U<n_f>}{2}+
\frac{A\Sigma_2(\omega)}{[1-B\Sigma_2(\omega)]}
\label{ansaz}
\end{equation}
where $\Sigma_2(\omega)$ is the second order self--energy defined in
terms of a modified bare local propagator
\begin{math}
\tg^{-1}={\cal G}^{-1}+\ep_f+\mu_0
\end{math}.
The parameter $\mu_0$ is adjusted so as to satisfy the 
Luttinger's theorem~\cite{luttwrd},
\begin{math}
\mbox{Im}\int_{-\infty}^0\frac{\partial\Sigma(\omega)}{\partial\omega}
G_f(\omega)d\omega = 0 \,. 
\end{math}
A and B are chosen so as to reproduce
the atomic limit and the high frequency behavior of the self-energy
at any filling, which yields~\cite{kajuet}
\begin{equation}
 A \equiv   \frac{n_f(2-n_f)}{n_0(2-n_0)}   ; \; 
 B  \equiv   \frac{4((1-n_f/2)U+\ep_f+\mu_0)}{n_0(2-n_0)U^2}  \,.
\end{equation}
Here $n_f  \equiv  2\int^{0}_{-\infty}\rho_f(\omega)d\omega$
is the f--band filling and $n_0$ is defined by
$n_0 \equiv 2\int^{0}_{-\infty}\rho_\tg(\omega)d\omega$,
where $\rho_f$ and $\rho_\tg$ are the spectral functions of $G_f$
and $\tg$ respectively: $\rho_f\equiv
-\frac{1}{\pi}\mbox{Im}G_f(\omega^+)$,
and $\rho_\tg\equiv -\frac{1}{\pi}\mbox{Im}\tg(\omega^+)$.


The conduction band filling is varied by varying $\ep_d$.
We maintain an $f$-band filling close to one, i.e $n_f\simeq1$ by
adjusting $\ep_f$. The actual value used for $n_f$ was $0.99$, 
and the accuracy achieved in fixing this value was $\sim 1$ in $10^5$.
For reasons of numerical convenience, we calculate the second--order 
self--energy directly for real frequencies
in two steps.  We first calculate the imaginary
part of $\Sigma_2$ using convolution integrals on a lorentzian
frequency grid; then use Kramer's-Kronig relations to find its real part.
Typically we achieve self--consistency of the Green functions within 3
to 4 iterations and the solution
for $\ep_f$ and $\mu_0$ is found within 10(outer loop) iterations using
a non--linear equation solving routine.

\section{Results}
We have performed the calculations described above for various values of
$U$ and $V$ and we present and discuss some representative results.
Since IPT is a perturbative technique, we have checked
its accuracy by comparing it with the earlier
QMC-MEM~\cite{niki} results for $U=1.5$ and $V=0.6$.

\begin{figure}[h]
\epsfxsize=6.5in
\epsffile{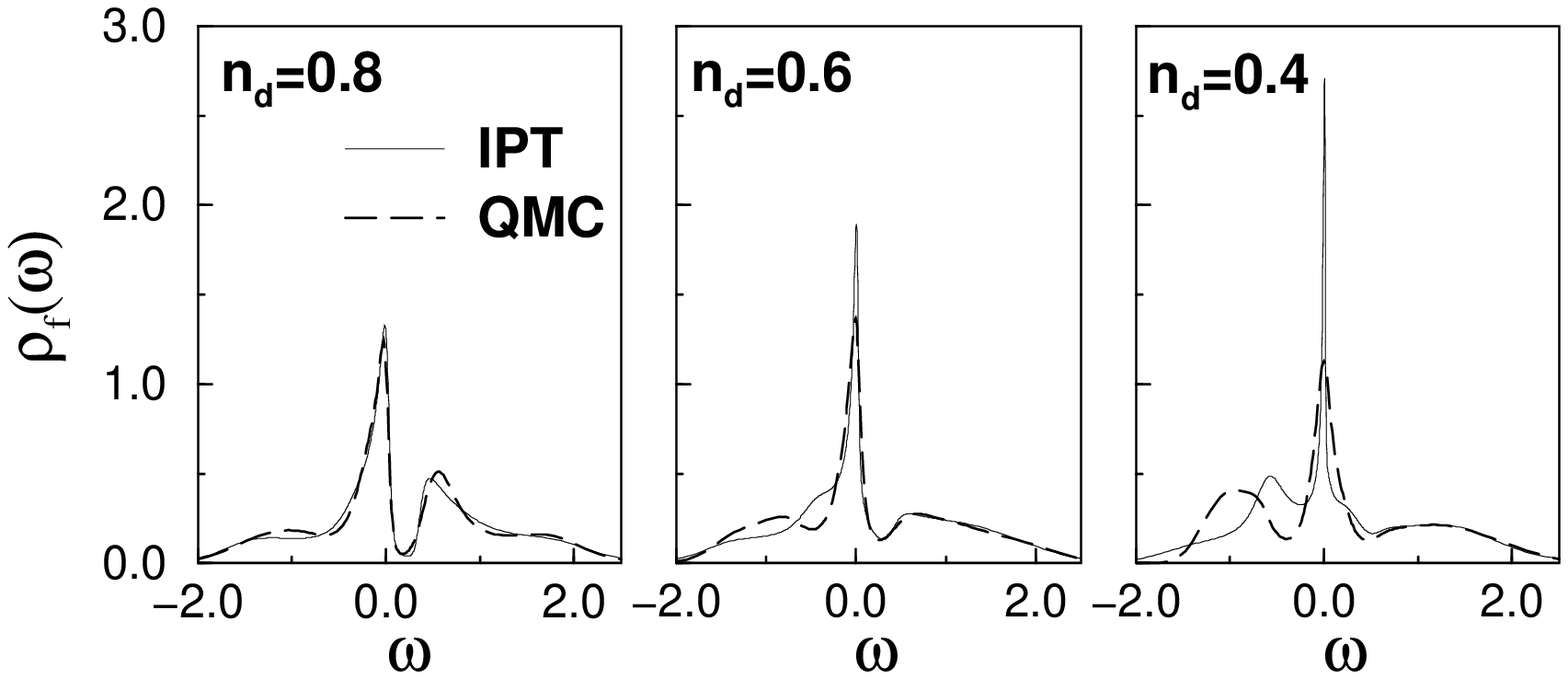}
\caption{ Comparison of the $f$ density of states obtained from IPT 
and QMC for three fillings of the conduction band, $n_d= 0.4, 0.6, 0.8,$ 
with $n_f\simeq 1$, for $U=1.5$, $V=0.6$. $T=0$ for IPT
and $T=0.025$ for QMC. The agreement is quite good for $d$-fillings
close to one.}
\label{fdos}
\end{figure}

Fig.~\ref{fdos} shows the $f$-spectral function for IPT at zero
temperature
and QMC-MEM at $T=0.025$ for three conduction band fillings namely
$n_d=0.4,0.6$ and $0.8$.  Both the QMC-MEM and IPT results share some
common features.  For small $n_d$ a single narrow Kondo peak is
centered at the Fermi energy $\omega=0$; however as $n_d\to 1$,
the peak broadens and splits in two, with both peaks shifted from
the Fermi energy.  Apparently, the latter is a remnant of the insulating
gap found when $n_f+n_d=2$. 
We see that the IPT results match rather well with QMC-MEM for $n_d=0.8$,
but as $n_d$ is decreased the deviation increases.
The difference in the width of the Kondo resonance between IPT and QMC
in the low $n_d$ or the exhaustion limit could be due to two
factors:$(1)$the calculations for the latter were carried out at higher 
temperatures which would lead to temperature broadening,$(2)$
 IPT is perturbative in $U$ while the impurity Kondo energy scale, $T_K$
is exponential in $U$.  Thus, IPT could become less accurate for very small
$n_d$ since the effective $\Gamma(0)$ is decreasing as $n_d$ decreases(see
below). 

\begin{figure}[h]
\epsfxsize=5.5in
\epsffile{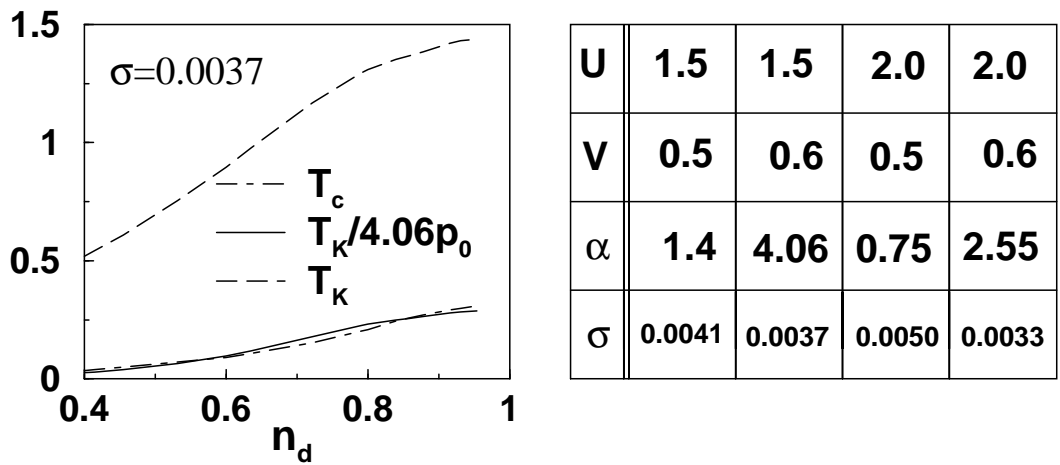}
\caption{ (a) The full width at half maximum for the Kondo resonance in
the $f$-spectral function ,(which is taken as a measure of ``$T_K$'' for
the SIAM and ``$T_c$'' for the PAM ) for $U=1.5$ and $V=0.6$ as a 
function of the conduction band filling. The solid line shows 
$T_K/\alpha(U,V)p_0$. $\sigma$ is the standard deviation for fitting the 
above relation to $T_c$ with $\alpha$ as the adjustable parameter.
(b)Table.I showing the values of $\alpha$ and $\sigma$ for four sets
of $U$ and $V$.}
\label{fwps}
\end{figure}

   In the QMC-MEM work~\cite{niki}, the SIAM Kondo scales
and the PAM coherence scales were identified from the $T\rightarrow 0$ limit
of the inverse of the appropriate impurity spin susceptibility, 
$\chi^{-1}_{imp}(T\rightarrow 0)$.
Since the IPT is not a conserving approximation, it does not provide 
a unique prescription for calculating $\chi_{imp}$. Hence, {\em as 
an alternate measure of the Kondo and coherence scales},
we calculate the Full Width at Half Maximum(FWHM) of the Kondo resonance
in the $f$--spectral function for the SIAM and PAM and identify these
with $T_K$ and $T_c$ respectively\footnote{\label{propo}For the SIAM, 
it is well known ~\cite{hewson} that both $\chi^{-1}_{imp}
(T\rightarrow 0)$ and the FWHM are proportional to the same $T_K$.}.
We present our results for these in Fig.~\ref{fwps}(a) for $U=1.5$ and $V=0.6$.
$T_c$ is seen to be much suppressed as compared to $T_K$, 
consistent with Nozi\e res' arguments~\cite{noz1,noz2} 
and the QMC--MEM results~\cite{niki}. 
As mentioned earlier, we find that $T_c$ and $T_K$ are related 
according to Eq.~\ref{Tckp0}. We have checked this 
for four sets of parameters and we show the values of $\alpha$ and 
$\sigma$(standard deviation of the fit in Eq.~\ref{Tckp0}) 
in Fig.~\ref{fwps}(b) for each of these sets.

The PAM, even at zero temperature, clearly has many other scales,
{\it eg.} $\kappa^{-1}=\left[\partial^2\Sigma_{\mbox{Im}}/\partial
\omega^2|_{\omega=0}\right]^{-1}$, although they are also related
to one other.  In the exhaustion limit, they are all typically 
suppressed compared to $T_K$, though to different degrees.
We have studied all of them, and hope to discuss them elsewhere.

But the important point is that, given that the PAM is being 
treated using DMFT, all the energy scales
in the problem are clearly determined by $U$ and by the effective 
hybridization of the self consistent Anderson--impurity problem
\begin{math}
\Gamma_{PAM}(\omega)\equiv\mbox{\bf Im}{\cal G}^{-1}=
\mbox{\bf Im}\left(G_{f,loc}^{-1}+\Sigma\right) \,.
\end{math}

\begin{figure}[h]
\epsfxsize=6.5in
\centerline{ \epsffile{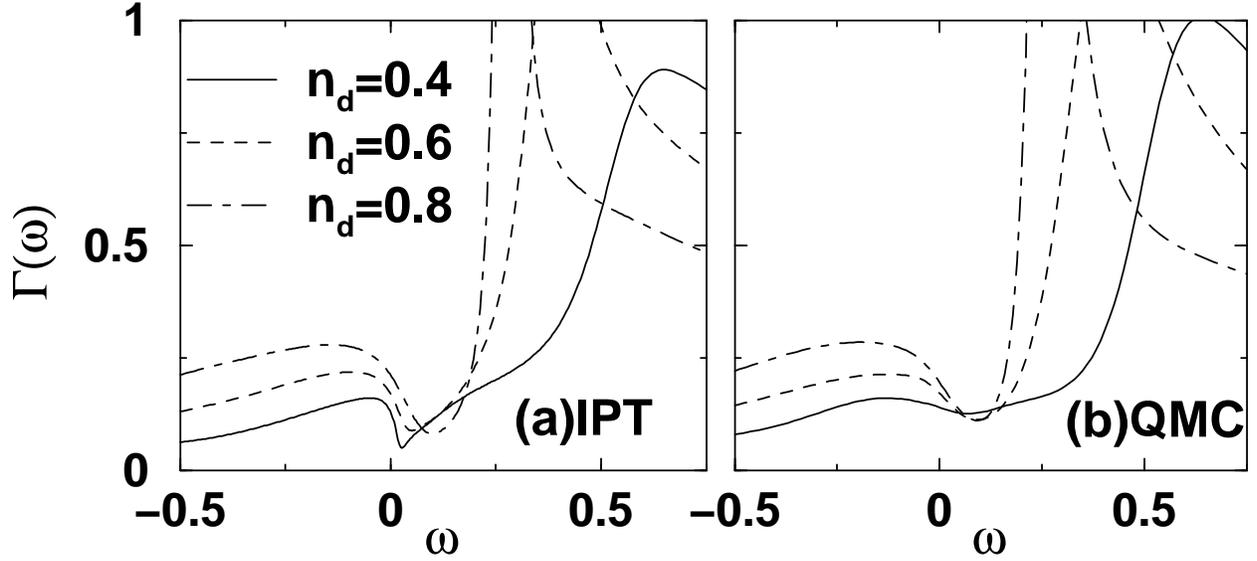}}
\caption{ The effective hybridization within (a) IPT and (b) QMC--MEM 
for the same parameters as Fig.~\ref{fdos} showing the decrease in
the number of states available for Kondo screening near the Fermi level.}
\label{tothyb}
\end{figure}

\begin{figure}[h]
\epsfxsize=6.5in
\centerline{ \epsffile{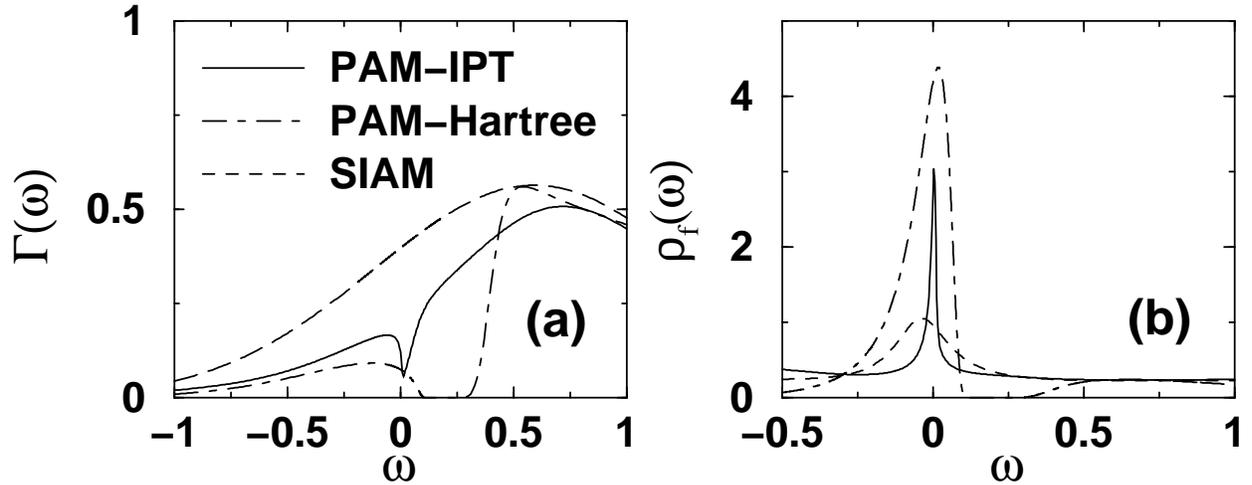}}
\caption{(a)The PAM--IPT ,the PAM--Hartree and the SIAM hybridization showing 
the effects of lattice coherence and exhaustion for $U=1.5$,$V=0.5$, 
and $n_d=0.4$. Near the Fermi level, the PAM--IPT $\Gamma$ has a dip
reflecting a strong energy dependence. Both the PAM--IPT and Hartree $\Gamma$
are seen to be much suppressed compared to the SIAM $\Gamma$ .
(b)The corresponding $f$--spectral functions 
are shown for the same parameters as (a). The Kondo
peak broadens for the three cases progressively.}
\label{hartree}
\end{figure}

$\Gamma_{PAM}$ is shown in Fig.~\ref{tothyb} as
calculated within QMC-MEM~\cite{niki} and IPT, and again the
agreement is reasonable over a fairly wide frequency scale.
In as much as ${\cal G}$, the (self--consistent) host propagator,
has built into it the self--energy, (and hence the ``Kondo'' screening)
processes at all other sites of the lattice except the one under 
consideration, {\em $\Gamma_{PAM}(\omega)$ necessarily encapsulates
the essential constraints of lattice coherence and collective screening
that leads to ``exhaustion''.} In Fig.~\ref{hartree}(a), we compare 
$\Gamma_{PAM}(\omega)$ with the bare hybridization $\Gamma_{SIAM}(\omega)
\equiv \pi V^2\rho(\omega +x)$( where $\rho$ and $x$ are defined below
Eq.~\ref{Tckp0}). Note that $\Gamma_{PAM}$ is overall strongly suppressed
compared to $\Gamma_{SIAM}$, and in addition has a sharp dip near
$\omega =0$. The suppression of $T_c$(and of other scales) compared to
$T_K$ is directly
related to the suppression of $\Gamma_{PAM}$ compared to $\Gamma_{SIAM}$,
as dramatically brought out by the direct comparison of their spectral
functions in Fig.~\ref{hartree}(b).

   In Fig.~\ref{hartree}(a), we also compare $\Gamma_{PAM}$ with 
the Hartree hybridization $\Gamma_{Hartree}(\omega)$($\equiv
\mbox{Im}(G^{-1}_{f,loc})$ where the latter is computed within the 
Hartree approximation). $\Gamma_{Hartree}$ can be interpreted as an
effective hybridization that arises when the self--consistent 
impurity problem is treated within the Hartree approximation.
Note that it is also strongly 
suppressed compared to $\Gamma_{SIAM}$, which can be thought of
as being due to exhaustion in the context of ``resonant level 
screening''. However, it is rather flat near the
fermi level and misses the strong energy dependence contained in the
dip in $\Gamma_{PAM}(\omega)$. As shown in Fig.~\ref{hartree}(b), the
FWHM of the Kondo resonance is of course much smaller than that of the
Hartree spectral function, since the latter has
only the resonant--level scales arising from $\Gamma_{Hartree}$, while 
the former includes(at least some of the) non--trivial correlation effects
which lead to local moment formation and {\em Kondo screening}
and the associated narrowing of scales. This suggests
that the strong {\em energy dependent} suppression of $\Gamma_{PAM}$
is the signature of exhaustion in the context of lattice coherent, 
collective, Kondo screening.

 From the arguments we have presented,
one would expect that there should be no such exhaustion effects
if $N_scr>>N_f$. We have studied such cases as well, and that this is 
indeed what we find.  
As we decrease $N_f$ keeping $n_d$ fixed, we find an increase in the 
FWHM of the resonance  at the femi level,
the dip in the effective hybridization
near the Fermi level gradually disapears, $\Gamma_{Hartree}$ and
$\Gamma_{PAM}$,  although still somewhat suppressed compared to 
$\Gamma_{SIAM}$, become almost identical and have a weak energy
dependence near the Fermi level, indicating that the screening processes
in this limit are predominantly "resonant level screening".

\section{Conclusions}
We have studied the ``exhaustion'' problem arising from lattice
coherent, collective, Kondo screening in the context
of the asymmetric PAM in the limit of infinite dimensions within IPT.
The IPT calculations, despite their limitations, capture( qualitatively
and in some regimes, even quantitatively) many of the features shown 
by the QMC calculations~\cite{niki}, including the
``exhaustion'' physics, which manifests itself as a strongly energy
dependent suppression of the effective hybridization $\Gamma(\omega)$.
{\em As a consequence}, all the (low) energy scales of the PAM are strongly
suppressed compared to the corresponding SIAM scales. We have presented 
detailed results for {\em one of them, namely the FWHM of the Kondo 
resonance}, and find that {\em its'} suppression ratio is roughly 
as proposed by Nozi\e res ( except for a $U,V$ dependent scale factor).
Together, these results also suggest that IPT would
be a useful method for incorporating moderate correlation
effects into {\em{ab initio}} calculations of Heavy Fermion 
materials~\cite{albers}.

It is a pleasure to acknowledge discussions with  A. Chattopadhyay,
J. Freericks, D.\ Hess, M. Hettler, D.\ Logan and Th.\ Pruschke. This 
work was supported by NSF grants DMR--9704021
and DMR--9357199, and the Ohio Board of
Regents Research Challenge Award(N. S. V and H. R. K).

\end{document}